\documentclass[sigconf]{acmart}
\AtBeginDocument{%
  }

\usepackage{booktabs}
\usepackage{multirow}
\usepackage{graphicx} % 如果需要调整表格宽度
\usepackage{algorithm}
\usepackage{bm}
\usepackage{algorithmic}
\usepackage{float}
\usepackage{amsmath}

\setcopyright{acmlicensed}
\copyrightyear{2018}
\acmYear{2018}
\acmDOI{XXXXXXX.XXXXXXX}
\acmConference[]{}{}{}

\acmISBN{978-1-4503-XXXX-X/2018/06}

\begin{document}

\title{Towards Universal Spatial Transcriptomics Super-Resolution: A Generalist Physically Consistent Flow Matching Framework}

\author{Xinlei Huang}
\email{huangxl_1999@163.com}
\affiliation{%
  \institution{Shenzhen People's Hospital}
  \city{Shenzhen}
  \state{Guangdong}
  \country{China}
}

\author{Weihao Dai}
\email{25B965035@stu.hit.edu.cn}
\affiliation{%
  \institution{Harbin Institute of Technology (Shenzhen)}
  \city{Shenzhen}
  \state{Guangdong}
  \country{China}
}

\author{Zijun Qin}
\email{Zijun.Qin1010@outlook.com }
\affiliation{%
  \institution{Great Bay University}
  \city{Dongguan}
  \state{Guangdong}
  \country{China}
} 

\author{Xin Yu}
\email{xinyu@stmail.ujs.edu.cn}
\affiliation{%
  \institution{Shenzhen People's Hospital}
  \city{Shenzhen}
  \state{Guangdong}
  \country{China}
}

\author{Di Wang}
\email{2300271071@email.szu.edu.cn}
\affiliation{%
  \institution{Shenzhen University}
  \city{Shenzhen}
  \state{Guangdong}
  \country{China}
}

\author{Yanran Liu}
\email{yanranliu7-c@my.cityu.edu.hk}
\affiliation{%
  \institution{Great Bay University}
  \city{Dongguan}
  \state{Guangdong}
  \country{China}
}

\author{Lixin Cheng}
\authornote{Both authors are corresponding authors.}
\email{easonlcheng@gmail.com}
\affiliation{%
  \institution{Shenzhen People's Hospital}
  \city{Shenzhen}
  \state{Guangdong}
  \country{China}
}

\author{Xubin Zheng}
\authornotemark[1]
\email{xbzheng@gbu.edu.cn}
\affiliation{%
  \institution{Great Bay University}
  \city{Dongguan}
  \state{Guangdong}
  \country{China}
}

\begin{abstract}
Spatial transcriptomics provides an unprecedented perspective for deciphering tissue spatial heterogeneity.
However, high-resolution spatial transcriptomic technology remains constrained by limited gene coverage, technical complexity, and high cost.
Existing spatial transcriptomics super-resolution methods from low resolution data suffer from two fundamental limitations: poor out-of-distribution generalization stemming from a neglect of inherent biological heterogeneity, and a lack of physical consistency.
To address these challenges, we propose SRast, a novel physically constrained generalist framework designed for robust spatial transcriptomics super-resolution. To tackle heterogeneity, SRast employs a strategic decoupling architecture that explicitly decouples gene semantics representation from spatial geometry deconvolution, utilizing self-supervised learning to align latent distributions and mitigate cross-sample shifts. Regarding physical priors, SRast reformulates the task as ratio prediction on the simplex, performing a flow matching model to learn optimal transport-based geometric transformations that strictly enforce local mass conservation. Extensive experiments across diverse species, tissues, and platforms demonstrate that SRast achieves state-of-the-art performance, exhibiting superior zero-shot generalization capabilities and ensuring physical consistency in recovering fine-grained biological structures.

\end{abstract}




% % 1. 隐藏 "ACM Reference Format" (引用格式)
% \settopmatter{printacmref=false}

% % 2. 隐藏 "Copyright" (版权声明)
% \setcopyright{none}
% \makeatletter
% \renewcommand\footnotetextcopyrightpermission[1]{}
% \makeatother



\maketitle

\section{Introduction}
The advent of spatial transcriptomics (ST) has fundamentally revolutionized our understanding of complex biological systems by providing an unprecedented lens to decipher tissue heterogeneity and cellular interactions at spatial resolution~\cite{staahl2016visualization,eng2019transcriptome,marx2021method}.
Spatial resolution is a critical factor governing ST data quality; although higher resolution facilitates the delineation of intricate sub-cellular structures, it comes at a prohibitive economic cost~\cite{moses2022museum,mao2024spatialqc}.
This inherent trade-off limits the scalability of high-resolution technologies to large-scale clinical cohorts, thereby hindering the excavation of universally applicable and fine-grained biological insights from low-cost datasets. To transcend this limitation, spatial transcriptomics super-resolution techniques have emerged as a cost-effective computational alternative.

Spatial transcriptomics super-resolution aims to generate high-resolution spatial gene expression profiles based on low-resolution inputs.
To this end, a variety of computational frameworks have been developed, ranging from statistical models to deep learning architectures.
BayesSpace~\cite{BayesSpace} introduces a Bayesian statistical framework to model latent gene expression clusters at the sub-spot level.
SpaVGN~\cite{SpaVGN} constructs a graph convolutional network~\cite{GCN} to enhance spatial smoothness and reconstruction quality.
iStar~\cite{iSTAR} integrates histology images as an auxiliary modality, leveraging high-frequency texture information to guide the super-resolution process.
iSCALE~\cite{iSCALE} and STRESS~\cite{STRESS} establish the mapping between low- and high-resolution data by employing advanced deep neural network architectures.
Although these methods perform reasonably well on in-distribution data, they tend to overfit the statistical features of specific samples. Moreover, their neglect of the inherent heterogeneity of biological data and the strict physical conservation laws governing biological entities severely compromises their generalization capability and reliability.

Biological data inherently exhibits severe heterogeneity arising from variations across species, individuals, tissues, and experimental batches~\cite{stuart2019integrative,tran2020benchmark}. Constrained by sequencing technologies and unavoidable experimental noise, spatial transcriptomics slices from different batches manifest significant sample heterogeneity~\cite{longo2021integrating}. Such heterogeneity induces distributional shifts, which cause catastrophic degradation in vanilla neural network models, particularly when deployed on the Out-of-Distribution (OOD) scenario~\cite{hendrycks2016baseline,koh2021wilds}.
Although existing methods attempt cross-sample validation, relying on adjacent slices from the same tissue is insufficient to rigorously benchmark this heterogeneity challenge. In practice, applying models to unseen tissue sections remains a formidable OOD obstacle. Furthermore, the spatial expression distributions of genes themselves exhibit marked heterogeneity across samples; the same gene may display distinctly different spatial trends in different tissue contexts. Existing methods couple gene semantic representations with spatial distribution reconstruction. This leads to an over-reliance on tissue-specific spatial patterns (i.e., spurious correlations), thereby further constraining the model's OOD generalization capability.

Another critical yet overlooked bottleneck lies in the problem formulation of the ST super-resolution task. Unlike traditional natural image super-resolution or generation tasks—which are typically cast as unbounded regression problems—ST super-resolution possesses an inherent physical prior: each Low-Resolution (LR) spot essentially functions as an aggregate "bulk" sample comprising multiple High-Resolution (HR) sub-spots.
This prior effectively recasts the task as a spatial deconvolution process, necessitating strict adherence to the physical constraint of local mass conservation—specifically, the sum of gene expression values in the predicted super-resolution (SR) sub-spots must equal the observed expression in the corresponding LR spot. However, by modeling this as an unbounded regression task, existing methods yield SR outputs where the aggregated sum significantly deviates from the LR observations. This discrepancy—a systematic failure we explicitly quantify in our experiments-is not only theoretically counter-intuitive but also introduces additional noise and hallucinations in practice.

To address these challenges, we propose SRast, a novel spatial transcriptomics super-resolution model. 
SRast employs an adaptive decoupling framework consisting of Structure-Aware Semantic Alignment (SASA) and Physically Constrained Flow Matching (PCFM).
SASA performs self-supervised training on the target dataset to characterize gene spatial patterns and align distributions across samples. 
PCFM employs a flow matching model to learn universal, optimal transport-based geometric rules from a large-scale, multi-species, and multi-tissue dataset. Crucially, unlike existing methods that treat super-resolution as an unbounded regression task, SRast reformulates the problem as ratio prediction on the simplex. Predicting gene allocation proportions via flow matching, it ensures adherence to local mass conservation. 
Extensive experiments across various species, tissues, and sequencing platforms demonstrate that SRast achieves state-of-the-art performance, exhibiting superior zero-shot generalization in recovering fine-grained biological structures while ensuring strict physical consistency.
Our contributions are summarised as follows:
\begin{itemize}
\item We propose SRast, a novel framework that addresses inherent biological heterogeneity to achieve superior zero-shot generalization across diverse species and tissues.
\item We decouple gene semantic representation from spatial geometry deconvolution, enabling the model to learn universal geometric rules while mitigating overfitting to sample-specific patterns.
\item We redefine the super-resolution task as ratio prediction on the simplex, utilizing flow matching to strictly enforce local mass conservation and ensure physical consistency.
\end{itemize}
\vspace{-0.2cm}

\section{Related works}
To bridge the resolution gap, spatial transcriptomics super-resolution (STSR) algorithms upscale low-resolution data to recover missing biological granularity, effectively bypassing the constraints of sequencing hardware. Existing STSR approaches can be broadly categorized into three classes: traditional interpolation, statistical methods, and deep learning-based approaches. Early attempts primarily utilized traditional interpolation techniques, such as Bilinear~\cite{bilinear}, Bicubic~\cite{bicubic}, and Gaussian~\cite{gaussian} interpolation, which estimate sub-spot values via mathematical averaging but often yield overly smoothed results. Subsequently, statistical frameworks were introduced to incorporate biological context; for instance, BayesSpace~\cite{BayesSpace} utilizes Markov Random Fields to model latent gene expression clusters at the sub-spot level.
Recent advancements have largely shifted towards deep learning architectures to capture complex non-linear mappings. SpaVGN~\cite{SpaVGN} constructs a graph convolutional network to enhance spatial smoothness and reconstruction quality by aggregating neighborhood information, while iStar~\cite{iSTAR} integrates histology images as an auxiliary modality to leverage high-frequency texture information. Similarly, methods like iSCALE~\cite{iSCALE} and STRESS~\cite{STRESS} employ advanced neural networks to extract non-linear features for establishing the mapping between low- and high-resolution data. However, these methods are fundamentally limited by their formulation as unbounded regression tasks, which violates the physical axiom of local mass conservation, leading to biological hallucinations, and their tendency to overfit sample-specific statistical features, resulting in poor generalization on Out-of-Distribution data.

\section{Method}
\begin{figure*}[t]
  \centering
  \includegraphics[ width=1.0\textwidth]{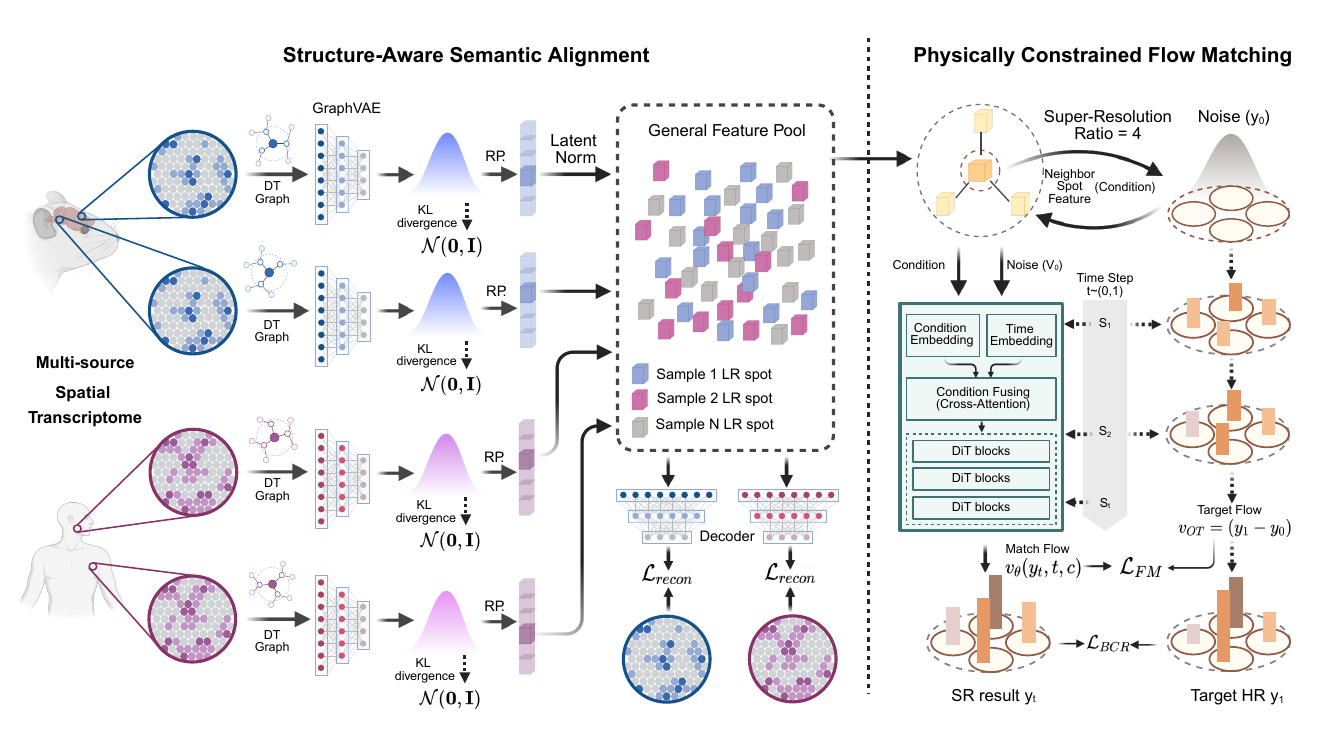}
  \caption{Overview of our proposed SRast framework. (Left) Structure-Aware Semantic Alignment: The model employs a sample-specific GraphVAE to generate sample representations based on the Dual-Topology (DT) Graph and utilizes Latent Norm to eliminate batch effects across multi-source data, constructing a unified General Feature Pool. (Right) Physically Constrained Flow Matching: Built upon the DiT architecture, the framework injects general features and their neighborhood features as conditions. It learns the optimal transport flow from noise to high-resolution ratios via flow matching and leverages a KL divergence constraint to ensure the generated distribution aligns with the target high-resolution distribution. 
  }
  \label{_frame}
\end{figure*}

\subsection{Preliminary}
Spatial transcriptomics (ST) super-resolution aims to computationally reconstruct high-fidelity, high-resolution (HR) gene expression profiles from low-resolution (LR), coarse-grained observations.
Formally, let $\mathcal{D}_{LR} = \{(\mathbf{X}_{LR}, \mathbf{C}_{LR})\}$ denote the observed LR dataset, where $\mathbf{X}_{LR} \in \mathbb{R}^{N \times G}$ represents the gene expression matrix for $N$ spots and $G$ genes, and $\mathbf{C}_{LR} \in \mathbb{R}^{N \times 2}$ denotes the corresponding spatial coordinates. The goal is to learn a mapping function $\mathcal{F}$ to predict the super-resolution expression $\mathbf{X}_{SR} \in \mathbb{R}^{M \times G}$, where each LR spot $i$ is spatially decomposed into a set of $K$ SR sub-spots $\mathcal{S}_i$ (i.e., $M = N \times K$):
\begin{equation}
\mathbf{X}_{SR} = \mathcal{F}(\mathbf{X}_{LR}, \mathbf{C}_{LR})
\end{equation}

Existing methods, typically modeled as unbounded regression tasks, are severely limited by the inherent biological heterogeneity and the variability of gene spatial expression patterns, and fail to generalize to Out-of-Distribution (OOD) challenges across different species, tissues, and sequencing platforms. To tackle this, SRast introduces a decoupled framework as illustrated in Figure 1.

\textbf{Structure-Aware Semantic Alignment (SASA)} focuses on learning robust gene representations and mitigating cross-sample distributional shifts via self-supervised learning on the target dataset. We define a structure-aware encoder $\Phi_{en}$ that maps LR data to a normalized, distribution-aligned latent space $\mathcal{Z}$:
\begin{equation}
\mathbf{Z} = \text{LatentNorm}(\Phi_{en}(\mathbf{X}_{LR}, \mathcal{G}_{LR}))
\end{equation}
where $\mathcal{G}_{LR}$ is the dual-topology graph  constructed from $\mathcal{D}_{LR}$, containing both local spatial neighborhoods and global gene representations. The goal is to obtain $\mathbf{Z}$ that is invariant to batch effects but sensitive to intrinsic biological semantics.

\textbf{Physically Constrained Flow Matching (PCFM).} We reformulate the super-resolution task as \textit{Ratio Prediction on the Simplex}. SRast learns a universal flow matching model based on the unified gene representations from SASA stage to capture Optimal Transport (OT) geometric laws. Specifically, SRast aims to learns a general flow trajectory via a continuous vector field $\mathbf{v}_\theta$ that pushes a prior noise distribution $p_0$ to the target ratio distribution on the simplex $\Delta^{K-1}$. The generative mapping $\mathcal{M}$ is defined as:
\begin{equation}
\mathbf{Y} = \mathcal{M}(\mathbf{Z}, \mathbf{C}_{HR}) = \mathbf{Y}_0 + \int_0^1 \mathbf{v}_\theta(\mathbf{Y}_t, t, \mathbf{Z}) dt
\end{equation}
where $\mathbf{Y}$ represents the unconstrained logits of the allocation ratios, t is the continuous time variable defined in the interval $[0, 1]$ that parameterizes the flow trajectory.
The final SR gene expression is obtained by projecting the predicted logits back to the simplex and strictly enforcing local mass conservation:
\begin{equation}
\hat{\mathbf{x}}_{j,g}^{SR} = \text{Softmax}_{k \in \mathcal{S}_i}(\mathbf{y}_{k,g}) \cdot x_{i,g}^{LR}, \quad \forall j \in \mathcal{S}_i
\end{equation}
where $\mathcal{S}_i$ denotes the set of $K$ sub-spots decomposed from LR spot $i$.
This ensures that the aggregated expression of the reconstructed sub-spots strictly equals the observed LR expression, guaranteeing physical consistency.

\subsection{Structure-Aware Semantic Alignment}

The intrinsic heterogeneity of biological samples, manifesting across distinct species, tissues, and batches, poses a formidable challenge for super-resolution generalization. Conventional models falter on Out-of-Distribution (OOD) data as they inadvertently conflate intrinsic gene semantics with sample-specific spatial distribution, effectively overfitting to spurious correlations. To alleviate this entanglement, SRast implements a decoupling framework. The primary objective of the SASA stage is to establish a unified, distribution-aligned semantic representation. By rigorously standardizing the latent distributions of diverse samples, we aim to mitigate domain-specific statistical shifts while preserving structure-aware gene semantics, thereby furnishing a stable, invariant condition for the subsequent geometric generation.

Formally, given the low-resolution gene expression $\mathbf{X}_{LR} \in \mathbb{R}^{N \times G}$ and spatial coordinates $\mathbf{C}_{LR} \in \mathbb{R}^{N \times 2}$, we first construct a dual-topology graph $\mathcal{G}=(\mathcal{V}, \mathcal{E})$ to capture both local spatial contexts and non-local semantic similarities. 
To define the edge set $\mathcal{E}$, we first reduce the dimensionality of the raw expression $\mathbf{X}_{LR}$ via Principal Component Analysis (PCA) to obtain dense semantic features $\mathbf{Z}_{PCA} \in \mathbb{R}^{N \times D}$. 
We then generate two distinct adjacency structures: a spatial neighbor graph $\mathcal{E}_{spa}$ constructed by identifying $k$-nearest neighbors based on Euclidean distances in the physical coordinate space $\mathbf{C}_{LR}$, and a semantic neighbor graph $\mathcal{E}_{sem}$ constructed by identifying k-nearest neighbors based on cosine similarity in the PCA feature space $\mathbf{Z}_{PCA}$. The final edge set is defined as the union of these two topologies, ensuring that the subsequent graph convolution aggregates information from both spatially adjacent and functionally similar spots:
\begin{equation}
\mathcal{E} = \mathcal{E}_{spa} \cup \mathcal{E}_{sem} = \text{KNN}_{Euclidean}(\mathbf{C}_{LR}, k) \cup \text{KNN}_{Cosine}(\mathbf{Z}_{PCA}, k)
\end{equation}

Based on this unified graph structure, we employ a Graph Variational Autoencoder (GVAE), implemented by a Graph Attention Networks~\cite{GAT} backbone, to model the spatial expression structure. For each spot $i$, the parameters of the posterior distribution $q_\phi(\mathbf{z}_i | \mathbf{x}_i, \mathcal{G})$ are inferred via:
\begin{equation}
\mathbf{h}_i = \text{GVAE}(\mathbf{x}_i, \mathcal{G})
\end{equation}
\begin{equation}
\boldsymbol{\mu}_i = \mathbf{W}_\mu \cdot \mathbf{h}_i, \quad \log\boldsymbol{\sigma}_i^2 = \mathbf{W}_\sigma \cdot \mathbf{h}_i
\end{equation}
where $\mathbf{W}\mu$ and $\mathbf{W}\sigma$ denote learnable projection matrices. The latent semantic vector $\mathbf{z}_i$ is subsequently sampled using the reparameterization trick~\cite{VAE}:
\begin{equation}\mathbf{z}_i = \boldsymbol{\mu}_i + \boldsymbol{\sigma}_i \odot \boldsymbol{\epsilon}, \quad \boldsymbol{\epsilon} \sim \mathcal{N}(\mathbf{0}, \mathbf{I})
\end{equation}

To further align feature distributions across varying batches, we introduce a Latent Normalization (LatentNorm) layer at the network bottleneck.
This layer standardizes the latent embeddings using running statistics ($\boldsymbol{\mu}_{e}, \boldsymbol{\sigma}^2_{e}$) accumulated via exponential moving average~\cite{ioffe2015batch} during the training phase: 
\begin{equation}
\hat{\mathbf{z}}_i = \boldsymbol{\gamma} \odot \frac{\mathbf{z}_i - \boldsymbol{\mu}_{e}}{\sqrt{\boldsymbol{\sigma}^2_{e} + \epsilon}} + \boldsymbol{\beta}
\end{equation}
where $\boldsymbol{\gamma}$ and $\boldsymbol{\beta}$ are affine parameters. 
Consequently, these normalized representation $\hat{\mathbf{z}}_i$ form a general feature pool, which serves as the input for PCFM stage.
This mechanism effectively reduces the first- and second-order distributional discrepancies between input samples, providing a canonicalized latent space that facilitates robust adaptation to unseen tissues.

To ensure fundamental reconstruction fidelity while maintaining a well-regularized latent space, we employ the standard Evidence Lower Bound (ELBO) loss to train encoders in the SASA stage. 
\begin{equation}\mathcal{L}_{ELBO} = |\mathbf{X}_{LR} - \text{Dec}(\hat{\mathbf{Z}})|_F^2 + \hat{w}_{KL} \mathcal{H}_{KL}(q_\phi(\mathbf{Z}|\mathbf{X}) | \mathcal{N}(\mathbf{0}, \mathbf{I}))
\end{equation}
where $\text{Dec}(\cdot)$ is sample-specific decoder, implemented by a Multi-layer perceptron.
$\mathcal{H}_{KL}$ denotes the KL divergence.
$\hat{w}_{KL}$ is a hyperparameter used to control the KL divergence weights, with a default value of 0.001.

\subsection{ Physically Constrained Flow Matching}
Existing spatial transcriptomics super-resolution methods predominantly model the task as an unbounded regression problem, denoted as $f: \mathbf{x}_{LR} \to \mathbf{x}_{SR}$. However, this formulation fundamentally ignores a basic physical constraint: \textit{local mass conservation}. Specifically, the total gene expression within a low-resolution (LR) region must strictly equal the sum of the expression in its constituent high-resolution (HR) sub-regions. Neglecting this constraint leads to physically impossible "hallucinations" and numerical inconsistencies in the generated data. To resolve this, we fundamentally redefine the problem from value regression to \textit{Ratio Prediction on the Simplex}. Instead of predicting arbitrary counts, we aim to learn how the known molecular mass of an LR spot is distributed onto the HR manifold. 

Formally, let $\mathbf{r}_{i,g} \in \Delta^{K-1}$ denote the allocation probability vector for gene $g$ in spot $i$.
A significant challenge in applying generative models to transcriptomic data is the extreme sparsity, where allocation ratios frequently approach zero ($r_{k,g} \to 0$). Direct logarithmic transformation of such values leads to numerical divergence.
To ensure numerical stability while mapping the simplex to an unconstrained Euclidean space, we introduce the Smoothed Centered Logits Transform (S-CLT). We apply a minimal smoothing factor $\epsilon_s$ (e.g., $10^{-6}$) to the probability vector before transformation. The robust mapping $\phi_{\epsilon_s}: \Delta^{K-1} \to \mathbb{R}^K$ is defined as:
\begin{equation}\mathbf{y}_{i,g} = \phi_{\epsilon_s}(\mathbf{r}_{i,g}) = \log(\mathbf{r}_{i,g} + \epsilon_s) - \frac{1}{K}\sum_{k=1}^K \log(r_{k,g} + \epsilon_s)\end{equation}

This regularization ensures that the logits remain bounded even for sparse genes, effectively preventing gradient explosion during training.
To rigorously justify the necessity of our Smoothed Centered Logits Transform (S-CLT) over traditional unbounded regression or naive log-ratio transformations, we analyze the numerical bounds and gradient behavior under extreme sparsity conditions common in spatial transcriptomics.

\textbf{Proposition 1 (Gradient Stability under Sparsity).} Consider a gene allocation ratio $r \in [0, 1]$. In naive regression targeting log-transformed space, the gradient norm diverges as $r \to 0$. In contrast, the SRast mapping $\phi_{\epsilon_s}$ guarantees $\mathcal{O}(\frac{1}{\epsilon_s})$-Lipschitz continuity, preventing numerical divergence.

\textbf{Proof.} Let $\mathcal{L}$ be a reconstruction loss defined on the transformed space. In standard log-space regression (often used to enforce positivity), the mapping is $f(r) = \log(r)$. The gradient with respect to the input ratio is:
\begin{equation}
\lim_{r \to 0^+} \left| \frac{\partial f}{\partial r} \right| = \lim_{r \to 0^+} \frac{1}{r} = \infty
\end{equation}

This singularity causes gradient explosion during backpropagation when the ground truth contains zeros (sparsity), a pervasive issue in ST data that we term "Zero-Inflation Barrier."
In SRast, the S-CLT mapping $\Phi_{\epsilon_s}(r) = \log(r + \epsilon_s) - \frac{1}{K}\sum \log(r_k + \epsilon_s)$ introduces a smoothing factor $\epsilon_s$. 
The partial derivative with respect to any component $r_j$ is bounded by:
\begin{equation}\left| \frac{\partial \Phi_{\epsilon_s}}{\partial r_j} \right| \leq \frac{1}{r_j + \epsilon_s} \leq \frac{1}{\epsilon_s}
\end{equation}

Since $\epsilon_s$ is a fixed constant (e.g., $10^{-6}$), the gradient is strictly bounded from above. This ensures that the vector field $v_{\theta}$ remains Lipschitz continuous even when transporting mass to effectively zero-probability regions on the simplex boundary.

\textbf{Proposition 2 (Boundedness of the Probability Flow).} The flow trajectory in the Euclidean space $\mathbb{R}^K$ is strictly confined within a hypercube, ensuring the generated flow never diverges to infinity.

\begin{table*}[th]
\fontsize{10}{10}\selectfont
\centering
\caption{Cross-sample zero-shot quantitative comparison of super-resolution performance on Human and Mouse datasets across 4$\times$ and 10$\times$ upscaling ratios. Best results are highlighted in bold. Underline identifies the second-best results. $\uparrow$ indicates higher is better, and $\downarrow$ indicates lower is better.}
\label{tab:performance_comparison}
\resizebox{\textwidth}{!}{
\begin{tabular}{llcccccccc}
\toprule
\multirow{2}{*}{\textbf{Class}} & \multirow{2}{*}{\textbf{Method}} & \multicolumn{4}{c}{\textbf{Human}} & \multicolumn{4}{c}{\textbf{Mouse}} \\
\cmidrule(lr){3-6} \cmidrule(lr){7-10}
 & & Spearman$\uparrow$ & PCC(gene)$\uparrow$ & FD$\downarrow$ & ARI$\uparrow$ & Spearman$\uparrow$ & PCC(gene)$\uparrow$ & FD$\downarrow$ & ARI$\uparrow$ \\
\midrule
\multicolumn{10}{c}{\textbf{SR Ratio: 4$\times$}} \\
\midrule
\multirow{3}{*}{Traditional} 
 & Bilinear & \underline{0.4721} & \underline{0.4100} & 2.8400 & 0.1629 & \underline{0.5575} & \underline{0.3977} & 2.7867 & 0.2005 \\
 & Bicubic & 0.3961 & 0.4126 & 2.9180 & 0.1764 & 0.3723 & 0.3973 & 2.8607 & 0.2149 \\
 & Gaussian & 0.3313 & 0.0713 & 3.2230 & 0.0419 & 0.3707 & 0.0807 & 3.1690 & 0.1875 \\
\midrule
\multirow{5}{*}{ST} 
 & STRESS & 0.3391 & 0.0628 & 1.9985 & 0.0943 & 0.4370 & 0.0632 & 2.2063 & 0.1899 \\
 & SpaVGN & 0.3212 & 0.1482 & 3.2730 & 0.0674 & 0.3926 & 0.0764 & 4.0647 & 0.0464 \\
 & BayesSpace & 0.4036 & 0.3255 & 3.1145 & 0.1581 & 0.5114 & 0.3551 & 3.0373 & 0.2460 \\
 & iStar & 0.3661 & 0.1906 & \underline{1.7054} & \underline{0.1861} & 0.3659 & 0.1879 & \underline{2.1993} & 0.2485 \\
 & iSCALE & 0.3700 & 0.1603 & 1.9335 & 0.0761 & 0.3689 & 0.0769 & 2.4360 & \underline{0.2854} \\
\midrule
\textbf{Our} & \textbf{SRast} & \textbf{0.6051} & \textbf{0.4880} & $\bm{4.00\times 10^{-8}}$ & \textbf{0.2002} & \textbf{0.7157} & \textbf{0.4900} & $\bm{4.59\times 10^{-8}}$ & \textbf{0.2926} \\
\midrule
\multicolumn{10}{c}{\textbf{SR Ratio: 10$\times$}} \\
\midrule
\multirow{3}{*}{Traditional} 
 & Bilinear & 0.3743 & 0.2438 & 8.2610 & 0.1396 & 0.4455 & 0.2396 & 8.0603 & 0.1998 \\
 & Bicubic & 0.3619 & 0.2475 & 8.3030 & \underline{0.1459} & 0.3502 & 0.2423 & 8.1363 & 0.2167 \\
 & Gaussian & 0.3549 & 0.0677 & 9.0645 & 0.0537 & 0.3705 & 0.0782 & 9.0427 & 0.1837 \\
\midrule
\multirow{5}{*}{ST} 
 & STRESS & 0.3535 & 0.0867 & 4.6150 & 0.1360 & 0.2195 & 0.0132 & 5.0183 & 0.1622 \\
 & SpaVGN & 0.2853 & 0.0881 & 8.5785 & 0.0476 & 0.4263 & 0.0690 & 8.8997 & 0.0625 \\
 & BayesSpace & \underline{0.3872} & \underline{0.2535} & 9.0895 & 0.1331 & \underline{0.4788} & \underline{0.2606} & 9.0349 & 0.2466 \\
 & iStar & 0.3630 & 0.1294 & \underline{4.4925} & 0.1426 & 0.3668 & 0.0787 & \underline{6.3363} & 0.2379 \\
 & iSCALE & 0.3674 & 0.1435 & 4.7355 & 0.0831 & 0.3683 & 0.0686 & 6.6410 & \underline{0.2538} \\
\midrule
\textbf{Our} & \textbf{SRast} & \textbf{0.4681} & \textbf{0.4478} & $\bm{6.00 \times 10^{-8}}$ & \textbf{0.1853} & \textbf{0.5875} & \textbf{0.2842} & $\bm{5.99 \times 10^{-8}}$ & \textbf{0.2886} \\
\bottomrule
\end{tabular}
}
\end{table*}

\textbf{Proof.}
For any valid probability distribution $r \in \Delta^{K-1}$, the mapped logit $y_{i,g}$ is bounded. The term $\log(r_{i,g}+\epsilon_s)$ lies in the interval $[\log(\epsilon_s), \log(1+\epsilon_s)]$. Consequently, the centered logit $y_{i,g}$ is bounded by the properties of the centered transformation:
\begin{equation}\| y_{i,g} \|_{\infty} \leq \max_{r \in \Delta^{K-1}} \left| \log(r+\epsilon_s) - \overline{\log(r+\epsilon_s)} \right| \leq \log\left(\frac{1+\epsilon_s}{\epsilon_s}\right)\end{equation}
This theoretically guarantees that our flow matching target $y_1$ and the interpolated path $y_t$ are always compact. Unlike unbounded regression where prediction $\hat{y}$ can drift to arbitrary magnitudes ($-\infty, +\infty$), SRast's flow is geometrically constrained to a safe zone isomorphic to the simplex interior.

Building upon the distribution-aligned, structure-aware latent representations $\hat{\mathbf{Z}} \in \mathbb{R}^{N \times D}$ generated in SASA stage, SRast employs a conditional flow matching framework to model the optimal transport flow from noise to the complex distribution of these ratios. Unlike diffusion models that rely on stochastic differential equations and require simulating complex noise schedules, SRast constructs a deterministic, optimal trajectory (Geodesic) from a prior noise distribution $p_0(\mathbf{y}) = \mathcal{N}(\mathbf{0}, \mathbf{I})$ to the data distribution $p_1(\mathbf{y})$. We define the probability density path $\psi_t(\mathbf{y}_0)$ as a linear interpolation:

\begin{equation}\psi_t(\mathbf{y}_0) = (1 - t)\mathbf{y}_0 + t \mathbf{y}_1, \quad t \in [0, 1]\end{equation}

This path induces a constant conditional vector field $\mathbf{u}_t$, which represents the straight-line velocity for transporting probability mass from the prior to the target:

\begin{equation}\mathbf{u}_t(\psi_t(\mathbf{y}_0) | \mathbf{y}_1, \mathbf{y}_0) = \frac{d}{dt}\psi_t(\mathbf{y}_0) = \mathbf{y}_1 - \mathbf{y}_0\end{equation}

We parameterize the time-dependent vector field using a neural network $\mathbf{v}_\theta(\mathbf{y}_t, t, \mathbf{c})$ based on the Diffusion Transformer (DiT)~\cite{peebles2023scalable} architecture. The network takes the interpolated state $\mathbf{y}_t \in \mathbb{R}^{M \times G}$ as input, projecting it into a hidden space $\mathbf{H}^{(0)}$. 
To effectively guide the generative process, we construct a composite conditioning signal that decouples temporal-semantic modulation from spatial encoding. For each HR spot $j$, the AdaLN modulation signal $\mathbf{c}_j$ fuses the sinusoidal time embedding $\mathbf{e}_t$ with the projected latent representation of its parent LR spot:
\begin{equation}
    \mathbf{c}_j = \text{MLP}_{fuse}\left( [\mathbf{e}_t \parallel \text{MLP}_{cond}(\mathbf{z}_{\pi(j)}^{LR})] \right)
\end{equation}
where $\pi(j) = i$ denotes the parent LR spot index for any given HR spot $j$ (i.e., $j \in \mathcal{S}_{\pi(j)}$).
Separately, positional information is incorporated by adding relative spatial embeddings and local index embeddings directly to the input hidden states:
\begin{equation}
    \mathbf{H}^{(0)} = \text{Proj}(\mathbf{y}_t) + \mathbf{e}_{pos}(\mathbf{c}_j^{HR} - \mathbf{c}_{\pi(j)}^{LR}) + \mathbf{e}_{idx}(j)
\end{equation}
where $\mathbf{e}_{idx}(j)$ is a learnable embedding encoding the local index of HR spot $j$ within its parent LR region. 
This decoupled design allows the AdaLN mechanism to focus on temporal-semantic modulation while spatial structure is preserved in the feature space.
This conditioning signal modulates the network dynamics via Adaptive Layer Normalization (AdaLN). 
Within each Transformer block, we employ AdaLN, which regresses scale, shift, and gate parameters from $\mathbf{c}_j$:
\begin{equation}
\text{AdaLN}(\mathbf{h}, \mathbf{c}) = (\mathbf{1} + \boldsymbol{\gamma}(\mathbf{c})) \odot \text{LayerNorm}(\mathbf{h}) + \boldsymbol{\beta}(\mathbf{c})
\end{equation}
where $(\boldsymbol{\gamma}, \boldsymbol{\beta}, \mathbf{g}) = \text{MLP}_{ada}(\mathbf{c})$, all initialized to zero. The gate $\mathbf{g}$ controls residual strength: $\mathbf{h}' = \mathbf{h} + \mathbf{g} \odot \text{SubLayer}(\text{AdaLN-Zero}(\mathbf{h}, \mathbf{c}))$.

The backbone consists of stacked DiT blocks that facilitate global information exchange via self-attention and cross-attention mechanisms. To further incorporate inductive bias for spatial smoothness, we append an SR Spatial Prior Module modeled as a graph neural layer. We construct a $k$-nearest neighbor graph $\mathcal{G}_{SR}$ based on physical coordinates and apply masked multi-head attention, restricting information flow to spatial neighbors. The final velocity estimate is produced via a gated fusion of the transformer features $\mathbf{H}^{(L)}$ and the spatially smoothed representations $\mathbf{H}_{spatial}$, followed by a linear projection with a residual skip connection:
\begin{equation} 
\mathbf{v}_\theta(\mathbf{y}_t, t, \mathbf{c}) = \mathbf{W}_{out} \left( (\mathbf{1} - \mathbf{g}) \odot \mathbf{H}^{(L)} + \mathbf{g} \odot \mathbf{H}_{spatial} \right) + \mathbf{W}_{skip}\mathbf{y}_t \end{equation}
where $\mathbf{g} = \sigma(\text{MLP}_{gate}([\mathbf{H}^{(L)} \parallel \mathbf{H}_{spatial}])) \in [0,1]$ is a learnable gate computed from the concatenation of transformer and spatial features. A higher gate value assigns greater weight to the spatially smoothed representations, enabling adaptive balance between global semantic consistency and local geometric fidelity.
The training objective is designed to guarantee that the learned vector field accurately reflects the optimal transport path while strictly adhering to simplex constraints. We employ the Flow Matching loss $\mathcal{L}_{FM}$ to minimize the mean squared error between the predicted velocity and the target conditional vector field:\begin{equation}\mathcal{L}_{FM} = \mathbb{E}_{t, \mathbf{y}_0, \mathbf{y}_1} \left[ | \mathbf{v}_\theta(\mathbf{y}_t, t, \mathbf{c}) - (\mathbf{y}_1 - \mathbf{y}_0) |_2^2 \right]
\end{equation}
However, regressing trajectories in unconstrained logit space can lead to numerical instability near the simplex boundaries where probabilities vanish. To mitigate this, we introduce a Boundary-Consistent Regularizer ($\mathcal{L}_{BCR}$). Instead of treating the velocity estimation purely as a regression problem, this term enforces a geometric constraint directly within the probability simplex. It requires that the terminal distribution implied by the current instantaneous velocity field—projected linearly to the end of the trajectory—must align with the ground-truth target distribution:
\begin{equation}
\mathcal{L}_{BCR} = w(t) \cdot \mathcal{H}_{KL}\left( \mathbf{p}_{target} | \mathbf{p}_{pred} \right)
\end{equation}
where $\mathbf{p}_{target} = \text{Softmax}(\mathbf{y}_1 / \tau)$ represents the ground-truth distribution on the simplex. $\mathbf{p}_{pred}$ denotes the predicted terminal distribution, derived by projecting the current state $\mathbf{y}_t$ along the learned vector field $\mathbf{v}_\theta$ to time $t=1$:
\begin{equation}
\mathbf{p}_{pred} = \text{Softmax}\left( \frac{\mathbf{y}_t + (1-t)\mathbf{v}_\theta(\mathbf{y}_t, t, \mathbf{c})}{\tau} \right)
\end{equation}

Here, $\hat{\mathbf{y}}_1 = \mathbf{y}_t + (1-t)\mathbf{v}_\theta$ represents the estimated destination of the flow based on the current trajectory slope. $\tau$ is the temperature parameter (default $\tau=2$)~\cite{hinton2015distilling}. The time-dependent weight $w(t) = t^2$ prioritizes fidelity as the flow approaches the target. By minimizing this forward KL divergence, the model is explicitly encouraged to respect the simplex geometry and enforce appropriate sparsity patterns, preventing "leakage" across boundaries.
The total loss is defined as 

\begin{equation}
\mathcal{L}_{total} = \mathcal{L}_{FM} + \mathcal{L}_{BCR}.
\end{equation}

During inference, we model the generation as an initial value problem~\cite{liu2022flow,lipman2022flow}. Starting from Gaussian noise $\mathbf{y}_0 \sim \mathcal{N}(\mathbf{0}, \mathbf{I})$, we integrate the learned ordinary differential equation using first-order Euler integration with a fixed number of steps $N_{steps}$ (default 50):
\begin{equation}
\mathbf{y}_{t+\Delta t} = \mathbf{y}_t + \mathbf{v}_\theta(\mathbf{y}_t, t, \mathbf{c}) \cdot \Delta t, \quad \Delta t = \frac{1}{N_{steps}}
\end{equation}

The final super-resolution gene expression is reconstructed by projecting the terminal logits $\hat{\mathbf{y}}_1$ back to the simplex via group-wise softmax and scaling by the observed total LR mass, ensuring intrinsic physical consistency and mass conservation.

\begin{equation} \hat{\mathbf{X}}_{SR,j} = \text{Softmax}_{k \in \mathcal{S}_{\pi(j)}}(\hat{\mathbf{y}}_1) \cdot \mathbf{x}_{\pi(j)}^{LR} \end{equation}

\section{Experiments}

\begin{table*}[tbp]
\fontsize{10}{10}\selectfont
\centering
\caption{Quantitative comparison of zero-shot cross-species generalization performance. Best results are highlighted in bold.
Underline identifies the second-best results.}
\label{tab:cross_species_comparison}
\resizebox{\textwidth}{!}{
\begin{tabular}{llcccccccc}
\toprule
\multirow{2}{*}{\textbf{Class}} & \multirow{2}{*}{\textbf{Method}} & \multicolumn{4}{c}{\textbf{Human $\rightarrow$ Mouse}} & \multicolumn{4}{c}{\textbf{Mouse $\rightarrow$ Human}} \\
\cmidrule(lr){3-6} \cmidrule(lr){7-10}
 & & \multicolumn{2}{c}{MISAR-seq} & \multicolumn{2}{c}{Stereo-seq} & \multicolumn{2}{c}{10x Visium} & \multicolumn{2}{c}{Stereo-seq} \\
\cmidrule(lr){3-4} \cmidrule(lr){5-6} \cmidrule(lr){7-8} \cmidrule(lr){9-10}
 & & Spearman & PCC(gene) & Spearman & PCC(gene) & Spearman & PCC(gene) & Spearman & PCC(gene) \\
\midrule
\multirow{3}{*}{Traditional} 
 & Bilinear & \underline{0.5575} & \underline{0.3977} & \underline{0.5951} & \underline{0.4165} & \underline{0.4721} & 0.4100 & \underline{0.5026} & \underline{0.4370} \\
 & Bicubic & 0.3723 & 0.3973 & 0.4218 & 0.4141 & 0.3961 & \underline{0.4126} & 0.2393 & 0.4067 \\
 & Gaussian & 0.3707 & 0.0807 & 0.4116 & 0.0352 & 0.3313 & 0.0713 & 0.2294 & 0.0274 \\
\midrule
\multirow{5}{*}{ST} 
 & STRESS & 0.4370 & 0.0632 & 0.5462 & 0.0443 & 0.3391 & 0.0628 & 0.4402 & 0.0377 \\
 & SpaVGN & 0.3926 & 0.0764 & 0.2925 & 0.0492 & 0.3212 & 0.1482 & 0.1121 & 0.0673 \\
 & BayesSpace & 0.5114 & 0.3551 & 0.2967 & 0.2436 & 0.4036 & 0.3255 & 0.1751 & 0.2566 \\
 & iStar & 0.3659 & 0.1879 & 0.4199 & 0.2710 & 0.3661 & 0.1906 & 0.2285 & 0.2978 \\
 & iSCALE & 0.3689 & 0.0769 & 0.4212 & 0.0827 & 0.3700 & 0.1603 & 0.2304 & 0.0326 \\
\midrule
\textbf{Our} & \textbf{SRast} & \textbf{0.7116} & \textbf{0.4534} & \textbf{0.6625} & \textbf{0.4591} & \textbf{0.6096} & \textbf{0.5029} & \textbf{0.5986} & \textbf{0.4899} \\
\bottomrule
\end{tabular}
}
\end{table*}

\begin{table}[htbp]
  \centering
  \caption{Quantitative comparison of zero-shot cross-platform generalization performance. Best results are highlighted in bold.
Underline identifies the second-best results.}
  \begin{tabular}{lcccc}
    \toprule
    \multirow{2}{*}{Platform} & \multicolumn{2}{c}{MISAR-seq (Mouse)} & \multicolumn{2}{c}{Stereo-seq (Human)} \\
    \cmidrule(lr){2-3} \cmidrule(lr){4-5}
    & Spearman & PCC(gene) & Spearman & PCC(gene) \\
    \midrule
    Bilinear     & \underline{0.5575} & \underline{0.3977} & \underline{0.5026} & \underline{0.4370} \\
    Bicubic      & 0.3723 & 0.3973 & 0.2393  & 0.4067  \\
    Gaussian     & 0.3707 & 0.0807 & 0.2294  & 0.0274  \\
    STRESS       & 0.4370 & 0.0632 & 0.4402  & 0.0377  \\
    SpaVGN       & 0.3926 & 0.0764 & 0.1121  & 0.0673  \\
    BayesSpace   & 0.5114 & 0.3551 & 0.1751  & 0.2566  \\
    istar   & 0.3659 & 0.1879 & 0.2285  & 0.2978  \\
    iSCALE  & 0.3689 & 0.0769 & 0.2304  & 0.0326  \\
    \textbf{SRast} & \textbf{0.7153} & \textbf{0.4788} & \textbf{0.5797} & \textbf{0.5770} \\
    \bottomrule
  \end{tabular}
  \vspace{-0.3cm}
\end{table}

\subsection{Experimental Setups}
\subsubsection{\textbf{Datasets.}}
We used 20 datasets across different tissue samples from humans and mice, containing 54 slides from 7 major spatial transcriptome sequencing platforms~\cite{long2024deciphering,liao2023integrated,zhang2023spatial,ben2023integration,yan2024mosaic,jiang2023simultaneous,cheng2022cellular,10x_breast_cancer_2022,10x_glioblastoma_2023,10x_tonsil_2023,10x_tonsil_he_2023,DLPFC1,DLPFC2,deepst,TNBC}.
The appendix provides dataset statistics and detailed descriptions.

\subsubsection{\textbf{Baselines.}}
We compare our work with traditional interpolation methods, including bilinear, bicubic, and Gaussian interpolation, and advanced spatial transcriptomics methods, including STRESS, SpaVGN, BayesSpace, iStar, iSCALE.

\subsubsection{\textbf{Metrics.}}
We evaluated the model performance on three levels: 1) super-resolution gene reconstruction quality, validated by Spearman coefficients and gene PCC; 2) physical consistency, validated by fractional deviation (FD); and 3) biological information preservation, using the recent state-of-the-art large-scale spatial transcriptome encoding method STAHD for encoding, and employing three clustering algorithms—K-means~\cite{macqueen1967multivariate}, Leiden~\cite{traag2019louvain}, and Mclust~\cite{scrucca2016mclust}—for clustering, calculating the evaluation metric Adjusted Rand Index (ARI) based on spatial domain labels.
The calculation details of the metric are formalized in the appendix.

\subsection{Cross-datasets zero-shot results}

We compared the cross-sample zero-shot generalization of SRast with baseline approaches. To ensure a rigorous evaluation, the entire test suite was excluded from the SRast training process. Since baselines typically lack robust generalization, they were trained on adjacent test slices to serve as a performance ceiling.
It should be noted that while SRast involves a rapid self-supervised adaptation on the test slices, this process is exclusively intended to extract latent LR features with a standard distribution. These features are then utilized for zero-shot super-resolution via the pre-trained flow matching model, which remains completely unexposed to any test set information during its training.

 The experimental results are presented in Table 1. SRast consistently achieves state-of-the-art performance across both 4x and 10x upscaling ratios. For instance, at a 4x ratio, SRast reaches a Spearman coefficient of 0.6051 for human data and 0.7157 for mouse data, significantly outperforming advanced models like BayesSpace and iStar. This performance gap stems from SRast’s ability to mitigate sample-specific distributional shifts. Furthermore, SRast exhibits exceptional physical consistency, with fractional deviations as low as $4.00 \times 10^{-8}$. This empirical result confirms that our formulation of the task as ratio prediction on the simplex effectively eliminates the numerical "hallucinations" and mass conservation violations prevalent in unbounded regression-based baselines.
Further qualitative visualization details of spatial gene expression are provided in the Appendix.

\vspace{-0.3cm}

\subsection{Cross-species zero-shot results}

We further evaluated the generalization capability of SRast through cross-species zero-shot testing. Specifically, the datasets were categorized into human and mouse origins. We then assessed the performance of SRast when trained on data from one species and tested on another. Various test scenarios involving different spatial resolutions were considered, including MISAR-seq, 10x Visium, and higher-resolution Stereo-seq. For the baseline methods, we continued to use adjacent slices from the test set for training to establish an empirical upper bound for their performance. 

Despite this extreme out-of-distribution (OOD) setting, Table 2 shows that SRast maintains its lead, consistently achieving state-of-the-art performance in this cross-species scenario. 
Specifically, in the Human $\rightarrow$ Mouse transition, SRast achieves a Spearman correlation of 0.7116 on MISAR-seq and 0.6625 on Stereo-seq, representing a significant improvement over the best-performing method. Similar trends are observed in the Mouse $\rightarrow$ Human experiments. On the 10x Visium platform, SRast reaches a Spearman correlation of 0.6096, whereas the second-best method (Bilinear) only achieves 0.4721. The performance gap is even more pronounced in high-resolution Stereo-seq data, where SRast (0.5986) surpasses the Bilinear baseline (0.5026) by a substantial margin. These results further validate SRast's superior zero-shot transferability and its potential for analyzing rare or emerging species where training data may be scarce.

\vspace{-0.2cm}

\subsection{Cross-platform zero-shot results}
To evaluate cross-platform transferability, we trained SRast on 10x Visium data and conducted zero-shot testing on MISAR-seq and Stereo-seq platforms. Baselines were trained on adjacent test slices to establish an empirical performance upper bound. As shown in Table 3, SRast consistently outperformed all competitors across both platforms; notably, on MISAR-seq, it achieved a Spearman correlation of 0.7153 and a PCC(gene) of 0.4788, significantly exceeding the best-performing baseline (Bilinear: 0.5575 and 0.3977).

The superiority of SRast is particularly evident in the Stereo-seq scenario, where it maintained a high PCC(gene) of 0.5770, while specialized ST methods like STRESS and BayesSpace exhibited sharp performance declines. These results underscore that SRast successfully captures platform-invariant spatial-biological representations. By surpassing in-distribution trained baselines through a pure zero-shot protocol, SRast demonstrates exceptional robustness to the domain shifts inherent in heterogeneous sequencing technologies.

\begin{table}[t]
\centering
\caption{Ablation study of SRast.}
% \vspace{-0.2cm}
\label{tab:ablation}
\resizebox{\columnwidth}{!}{% % 关键：将表格宽度缩放到单栏宽度
\begin{tabular}{lcccc}
\toprule
\multirow{2}{*}{\textbf{Configuration}} & \multicolumn{2}{c}{\textbf{Human}} & \multicolumn{2}{c}{\textbf{Mouse}} \\ 
\cmidrule(lr){2-3} \cmidrule(lr){4-5}
 & Spearman & PCC(gene) & Spearman & PCC(gene) \\ \midrule
\textbf{SRast (Full)} & \textbf{0.6051} & \textbf{0.4880} & \textbf{0.7157} & \textbf{0.4900} \\ \midrule
w/o SASA stage (PCA instead) & 0.5829 & 0.3745 & 0.7050 & 0.3810 \\ 
w/o Latent Norm & 0.5970 & 0.4566 & 0.7116 & 0.4534 \\ 
w/o $\mathcal{L}_{BCR}$ & 0.5916 & 0.3939 & 0.7085 & 0.3923 \\
w/o $\mathcal{L}_{FM}$ & 0.5953 & 0.4480 & 0.7109 & 0.4462 \\ 
\bottomrule
\end{tabular}%
}
% \vspace{-0.3cm}
\end{table}

\begin{figure}[t]
  \centering
  \includegraphics[ width=0.5\textwidth]{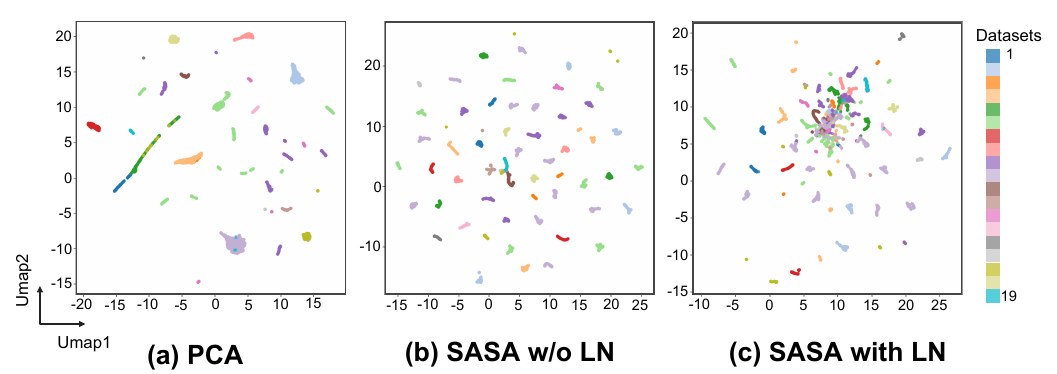}
  \caption{Umap visualization comparison of PCA features (a), SASA stage encoded features (b), and features after Latent Norm (c) across multiple datasets.
  % \vspace{-0.3cm}
}
  \label{_frame}
\end{figure}

\subsection{Ablation Studies}
To investigate the contribution of each component to SRast’s robust generalization, we conducted comprehensive ablation studies (Table 4) and visualized the latent space transformations (Figure 2). 
The removal of SASA stage (replacing GVAE with PCA) or the exclusion of LatentNorm leads to a significant performance decline. For instance, without LatentNorm, the Spearman correlation on human data drops from 0.6051 to 0.4566. This is further elucidated by the UMAP visualizations in Figure 2. While raw PCA and SASA stage without LatentNorm exhibit fragmented clusters and severe distribution shifts across multi-source data (Fig. 2a-b), the inclusion of LatentNorm successfully aligns these diverse samples into a canonicalized latent space (Fig. 2c). This alignment is crucial for zero-shot tasks, as it ensures that the PCFM stage model receives stable, species-invariant semantic conditions.
The ablation of Flow Matching loss ($\mathcal{L}_{FM}$) and the Boundary-Consistent Regularizer ($\mathcal{L}_{BCR}$) confirms their roles in geometric deconvolution. Interestingly, the model maintains reasonable performance even without explicit $\mathcal{L}_{FM}$, as the $\mathcal{L}_{BCR}$ provides a compensatory effect by guiding the flow towards the target distribution on the simplex.
Additionally, we verified the inference stability of our framework under different numerical solvers, as detailed in the Appendix.

\subsection{Inference Time Cost}
We evaluated the practical scalability of SRast by benchmarking its inference latency across varying data volumes. As illustrated in Figure 3, SRast’s inference time exhibits a strict linear growth pattern relative to the number of low-resolution (LR) spots. Notably, SRast demonstrates superior computational efficiency, with inference costs remaining lower than traditional interpolation baselines such as Bilinear and Bicubic interpolation. This predictable scaling behavior underscores its suitability for high-throughput clinical applications and large-scale spatial transcriptomics studies.

\begin{figure}[tb]
  \centering
  % \vspace{-0.2cm}
  \includegraphics[ width=0.5\textwidth]{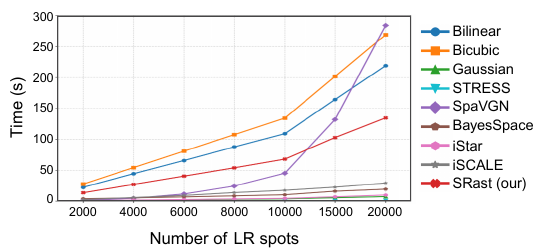}
  % \vspace{-0.5cm}
  \caption{Comparison of Running Time Cost between SRast and Baseline Methods as the Number of Inference Samples Increases
  }
  \vspace{-0.3cm}
  \label{_frame}
\end{figure}

\section{Limitations and Ethical Considerations}
This study exclusively utilizes publicly available, anonymized datasets. No new human or animal specimens were collected. The research strictly adheres to ACM ethical guidelines for the secondary analysis of existing data.

\section{Conclusion}
In this paper, we focus on the bottlenecks in spatial transcriptomics super-resolution arising from biological data heterogeneity and task definitions that violate physical principles, and introduce a generalist and physically consistent flow matching framework——SRast. To mitigate cross-sample and cross-species heterogeneity, SRast utilizes a structure-aware semantic alignment module integrated with a Latent Normalization layer to harmonize latent distributions across diverse datasets, providing a stable condition for generative reconstruction. Furthermore, we fundamentally redefine the super-resolution task as Ratio Prediction on the Simplex, employing a conditional Flow Matching model to learn optimal transport trajectories that strictly enforce the biological axiom of local mass conservation. Extensive experiments across diverse species, tissues, and sequencing platforms demonstrate that SRast achieves state-of-the-art performance, exhibiting superior zero-shot generalization and linear scalability for recovering fine-grained biological insights from low-resolution spatial datasets.

\bibliographystyle{ACM-Reference-Format}
\bibliography{Ref}

% \appendix
% \input{_Appendix}

\end{document}